\documentclass[twocolumn]{aastex631}

\usepackage{tabularx}
\usepackage{multirow}
\usepackage{array}

\newcolumntype{x}[1]{>{\centering\arraybackslash\hspace{0pt}}p{#1}}

\newcommand\tess{TESS}

\newcommand\logg{log~\textit{g}$_\star$}
\newcommand{\unit}[1]{\ensuremath{\, \mathrm{#1}}} 
\newcommand\earthmass{$M_{\oplus}$}
\newcommand\earthradius{$R_{\oplus}$}
\newcommand\solmass{$M_{\odot}$}

\usepackage{subfigure}
\usepackage{wrapfig}

\definecolor{lightblue}{rgb}{0.145,0.6666,1} 

\received{August 30, 2024}
\revised{November 04, 2024}
\accepted{December 05, 2024}
\submitjournal{AJ}

\begin{document}

\title{\textit{Searching for GEMS:} TOI-5688~A~b, a low-density giant orbiting a high-metallicity early M-dwarf\footnote{Based on observations obtained with the Hobby-Eberly Telescope (HET), which is a joint project of the University of Texas at Austin, the Pennsylvania State University, Ludwig-Maximillians-Universitaet Muenchen, and Georg-August Universitaet Goettingen. The HET is named in honor of its principal benefactors, William P. Hobby and Robert E. Eberly}}

\author[0009-0006-7298-619X]{Varghese Reji}
\affiliation{Department of Astronomy and Astrophysics, Tata Institute of Fundamental Research, Homi Bhabha Road, Colaba, Mumbai 400005, India}

\author[0000-0001-8401-4300]{Shubham Kanodia}
\affiliation{Earth and Planets Laboratory, Carnegie Science, 5241 Broad Branch Road, NW, Washington, DC 20015, USA}

\author[0000-0001-8720-5612]{Joe P. Ninan}
\affiliation{Department of Astronomy and Astrophysics, Tata Institute of Fundamental Research, Homi Bhabha Road, Colaba, Mumbai 400005, India}


\author[0000-0003-4835-0619]{Caleb I. Cañas}
\affiliation{NASA Postdoctoral Fellow}
\affiliation{NASA Goddard Space Flight Center, 8800 Greenbelt Road, Greenbelt, MD 20771, USA}

\author[0000-0002-2990-7613]{Jessica Libby-Roberts}
\affiliation{Department of Astronomy \& Astrophysics, 525 Davey Laboratory, The Pennsylvania State University, University Park, PA, 16802, USA}
\affiliation{Center for Exoplanets and Habitable Worlds, 525 Davey Laboratory, The Pennsylvania State University, University Park, PA, 16802}

\author[0000-0002-9082-6337]{Andrea S.J. Lin}
\affiliation{Department of Astronomy \& Astrophysics, 525 Davey Laboratory, The Pennsylvania State University, University Park, PA, 16802, USA}
\affiliation{Center for Exoplanets and Habitable Worlds, 525 Davey Laboratory, The Pennsylvania State University, University Park, PA, 16802}

\author[0000-0002-5463-9980]{Arvind F. Gupta}
\affiliation{NSF NOIRLab, 950 N. Cherry Ave., Tucson, AZ 85719, USA}

\author[0000-0002-5817-202X]{Tera N. Swaby}
\affiliation{Department of Physics \& Astronomy, University of Wyoming, Laramie, WY 82070, USA}
\author[0000-0002-2401-8411]{Alexander Larsen}
\affiliation{Department of Physics \& Astronomy, University of Wyoming, Laramie, WY 82070, USA}
\author[0000-0002-4475-4176]{Henry A. Kobulnicky}
\affiliation{Department of Physics \& Astronomy, University of Wyoming, Laramie, WY 82070, USA}

\author{Philip I. Choi}
\affiliation{Pomona College, 333 N. College Way Claremont, CA 91711}
\author{Nez Evans}
\affiliation{Pomona College, 333 N. College Way Claremont, CA 91711}
\author[0000-0002-8623-8268]{Sage Santomenna}
\affiliation{Pomona College, 333 N. College Way Claremont, CA 91711}
\author{Isabelle Winnick}
\affiliation{Pomona College, 333 N. College Way Claremont, CA 91711}
\author{Larry Yu}
\affiliation{Pomona College, 333 N. College Way Claremont, CA 91711}

\author[0000-0003-0353-9741]{Jaime A. Alvarado-Montes}
\affiliation{School of Mathematical and Physical Sciences, Macquarie University, Balaclava Road, North Ryde, NSW 2109, Australia}
\affiliation{The Macquarie University Astrophysics and Space Technologies Research Centre, Macquarie University, Balaclava Road, North Ryde, NSW 2109, Australia}

\author[0000-0003-4384-7220]{Chad F Bender}
\affiliation{Steward Observatory, University of Arizona, 933 N. Cherry Ave, Tucson, AZ 85721, USA}

\author[0000-0002-8035-1032]{Lia Marta Bernabò}
\affiliation{Institute of Planetary Research, German Aerospace Center (DLR), Rutherfordstrasse 2, 12489 Berlin}
\affiliation{McDonald Observatory and Center for Planetary Systems Habitability, The University of Texas at Austin, Austin, TX 78730, USA}

\author[0000-0002-6096-1749]{Cullen H. Blake}
\affiliation{Department of Physics and Astronomy, University of Pennsylvania, 209 S 33rd Street, Philadelphia, PA 19104, USA}

\author[0000-0001-9662-3496]{William D. Cochran}
\affiliation{McDonald Observatory and Center for Planetary Systems Habitability, The University of Texas at Austin, Austin, TX 78730, USA}

\author[0000-0002-2144-0764]{Scott A. Diddams}
\affiliation{Electrical, Computer \& Energy Engineering, University of Colorado, 425 UCB, Boulder, CO 80309, USA}
\affiliation{Department of Physics, University of Colorado, 2000 Colorado Avenue, Boulder, CO 80309, USA}

\author[0000-0003-1312-9391]{Samuel Halverson}
\affiliation{Jet Propulsion Laboratory, California Institute of Technology, 4800 Oak Grove Drive, Pasadena, California 91109}

\author[0000-0002-7127-7643]{Te Han}
\affiliation{Department of Physics \& Astronomy, The University of California, Irvine, Irvine, CA 92697, USA}

\author[0000-0002-1664-3102]{Fred Hearty}
\affiliation{Department of Astronomy \& Astrophysics, 525 Davey Laboratory, The Pennsylvania State University, University Park, PA, 16802, USA}
\affiliation{Center for Exoplanets and Habitable Worlds, 525 Davey Laboratory, The Pennsylvania State University, University Park, PA, 16802}

\author[0000-0002-9632-9382]{Sarah E. Logsdon}
\affiliation{NSF NOIRLab, 950 N. Cherry Ave., Tucson, AZ 85719, USA}

\author[0000-0001-9596-7983]{Suvrath Mahadevan}
\affiliation{Department of Astronomy \& Astrophysics, 525 Davey Laboratory, The Pennsylvania State University, University Park, PA, 16802, USA}
\affiliation{Center for Exoplanets and Habitable Worlds, 525 Davey Laboratory, The Pennsylvania State University, University Park, PA, 16802}

\author{Michael W. McElwain}
\affiliation{NASA Goddard Space Flight Center, 8800 Greenbelt Road, Greenbelt, MD 20771, USA}

\author[0000-0002-0048-2586]{Andrew Monson}
\affiliation{Steward Observatory, University of Arizona, 933 N. Cherry Ave, Tucson, AZ 85721, USA}

\author[0000-0003-0149-9678]{Paul Robertson}
\affiliation{Department of Physics \& Astronomy, The University of California, Irvine, Irvine, CA 92697, USA}

\author[0000-0001-9312-3816]{Devendra K Ojha}
\affiliation{Department of Astronomy and Astrophysics, Tata Institute of Fundamental Research, Homi Bhabha Road, Colaba, Mumbai 400005, India}

\author[0000-0001-8127-5775]{Arpita Roy}
\affiliation{Astrophysics \& Space Institute, Schmidt Sciences, New York, NY 10011, USA}

\author[0000-0002-4046-987X]{Christian Schwab}
\affiliation{School of Mathematical and Physical Sciences, Macquarie University, Balaclava Road, North Ryde, NSW 2109, Australia}

\author[0000-0001-7409-5688]{Gudmundur Stefansson}
\affiliation{Anton Pannekoek Institute for Astronomy, University of Amsterdam, Science Park 904, 1098 XH Amsterdam, The Netherlands}

\author[0000-0001-6160-5888]{Jason Wright}
\affiliation{Department of Astronomy \& Astrophysics, 525 Davey Laboratory, The Pennsylvania State University, University Park, PA, 16802, USA}
\affiliation{Center for Exoplanets and Habitable Worlds, 525 Davey Laboratory, The Pennsylvania State University, University Park, PA, 16802}
\affiliation{Penn State Extraterrestrial Intelligence Center, 525 Davey Laboratory, The Pennsylvania State University, University Park, PA, 16802, USA}

\correspondingauthor{Varghese Reji}
\email{varghesereji0007@gmail.com}

\begin{abstract}

We present the discovery of a low-density planet {orbiting the high-metallicity early M-dwarf TOI-5688 A b}. This planet was characterized as part of the search for transiting giant planets ($R \gtrsim8$ \earthradius{}) through the  \textit{Searching for GEMS} (Giant Exoplanets around M-dwarf Stars) survey.  The planet was discovered with the Transiting Exoplanet Survey Satellite (TESS), and characterized with ground-based transits from Red Buttes Observatory (RBO), the Table Mountain Observatory of Pomona College, and radial velocity (RV) measurements with the Habitable-Zone Planet Finder (HPF) on the 10 m Hobby Eberly Telescope (HET) and NEID on the WIYN 3.5 m telescope. From the joint fit of transit and RV data,{ we measure a planetary mass and radius of }$124\pm24$ M$_\oplus$ {($0.39\pm0.07$ M\textsubscript{J})} and $10.4\pm0.7$  R$_\oplus$ {($0.92\pm0.06$ R\textsubscript{J})} respectively. The spectroscopic and photometric analysis of the host star TOI-5688 A shows that it is a metal-rich ([Fe/H] $ = 0.47\pm0.16$ dex) M2V star, favoring the core-accretion formation pathway as the likely formation scenario for this planet. Additionally, Gaia astrometry suggests the presence of a wide-separation binary companion, TOI-5688 B, which has a projected separation of $\sim5"$ (1110 AU) and is an M4V, making TOI-5688 A b part of the growing number of GEMS in wide-separation binary systems.  
\end{abstract}


\section{Introduction}
\label{sec:intro}
M dwarfs are the most prevalent stars {in our Galaxy} \citep{2006AJ....132.2360H,2021AA...650A.201R} and tend to host more planets on average compared to FGK type stars \citep{2015ApJ...798..112M}. {Models of planet formation -- core-accretion \citep{POLLACK199662} and gravitational instability \citep{2006ApJ...643..501B} -- fail to explain the \textit{in-situ} formation of GEMS (Giant Exoplanets around M-dwarf Stars) because of the low mass of the host stars and their protoplanetary disks, lower surface density of dust and also longer formation timescales \citep{2004ApJ...612L..73L, 2005ApJ...626.1045I, 2006ApJ...648..666R}. Instead, recent studies try to explain the \textit{ex-situ} formation of GEMS through core accretion \citep{motivation} and gravitational instability \citep{2023ApJ...956....4B}, followed by migration to the current location. The  \textit{Searching for GEMS} survey aims to discover and characterize more GEMS to constrain current planet formation models empirically.}



Despite their rarity, the all-sky survey with the Transiting Exoplanet Survey Satellite \citep[TESS;][]{lens.org/003-780-753-054-02X} has discovered $\sim 25$ GEMS \citep{motivation} with precise mass measurements. \cite{Bryant_2023} reported that the occurrence rate of giant planets ($0.6 ~\text{R}_J \leq R_p \leq  2.0 ~\text{R}_J$) around a sample of $\sim 90,000$ low-mass stars ($\leq 0.71$ \solmass) observed with TESS is only $0.194 \pm 0.072\%$. A similar study of early M-dwarfs ($0.45\leq\text{M}_\star\leq0.65$ \solmass{}) conducted by \cite{2023AJ....165...17G} finds a consistent occurrence  rate for periods $0.8\leq \text{P} \leq 10$ days and radii of 7\earthradius{} $\leq$ R\textsubscript{p} $\leq$ 2 R\textsubscript{J} to be $0.27\pm0.09$\%.


In this manuscript, we describe the discovery of the transiting low-density planet TOI-5688 A b, using a combination of photometry from seven sectors of TESS, ground-based photometry on the 0.6 m telescope at Red Buttes Observatory and Pomona College 1 m telescope at NASA JPL's Table Mountain Facility, and also spectra and precise radial velocities (RVs) from the Habitable-zone Planet Finder (HPF) spectrograph and NEID spectrograph. {The system is cataloged by \cite{el-bardy2021} as a wide separation binary based on Gaia astrometry. In this manuscript, the host star is called TOI-5688 A, and the companion star of the binary system is called TOI-5688 B.} Section \ref{sec:observations} contains a description of the observations. Section \ref{sec:stellar} describes the estimation of stellar parameters of the host star and its binary companion. Section \ref{sec:joint} explains the joint fitting of transit and RV data {to obtain planetary parameters}. {We discuss our findings in} Section \ref{sec:discussion}, {including }a comparison with other planets hosted by M-dwarfs, {their formation mechanisms, as well as }wide-separation binary systems hosting GEMS. {Finally, we include a summary in }Section \ref{sec:summary}.

\section{Observations}\label{sec:observations}
\begin{table*}[b]
    \centering
    \caption{Summary of space-based and ground-based transit observations of TOI-5688 A b}
    \label{tab:tess_sect}
    \begin{tabular}{cccccc}
    \hline
    \hline
         Instrument & Date & Exposure time & Filter & Median PSF \\
         & UTC & (s) & & FWHM (") \\
         \hline
         TESS/S25& 2020 May 13 - 2020 June 08 & 1800 & T & 39.5 \\
         TESS/S26& 2020 June 08 - 2020 July 04 & 1800 & T & 39.5 \\
         TESS/S40 & 2021 June 24 - 2021 July 23 & 600 & T & 39.5 \\
        TESS/S51 & 2022 April 22 - 2022 May 18 & 600 & T & 39.5\\
        TESS/S52 & 2022 May 18 - 2022 June 13 & 600 & T & 39.5\\
        TESS/S53 & 2022 June 13 - 2022 July 09 & 600 & T & 39.5\\
        TESS/S54 & 2022 July 09 - 2022 August 09 & 600 & T & 39.5\\
        0.6 m RBO & 2022 October 09 & 240 & Bessel I & 1.49 \\
        0.6 m RBO & 2023 May 15 & 240 & Bessel I & 1.52 \\
        1.0 m TMF & 2023 July 19 & 10 & SDSS \textit{i'} & 4.6\\
        \hline
    \end{tabular}
    
\end{table*}

\subsection{TESS}\label{sec:TESS}
TESS \citep{lens.org/003-780-753-054-02X} observed TOI-5688 A (TIC 193634953, 2MASS J17474153+4742171, APASS 53641204, Gaia DR3 1363205856494897024) over seven sectors (25, 26, 40, 51, 52, 53 and 54). The planet candidate was{ identified in the TESS Faint Star Search \citep{Kunimoto_2022}} with an orbital period of $\sim2.95$ days.
We extracted aperture photometry flux from TESS full-frame images using \texttt{eleanor} \citep{eleanor}, which uses TESScut \citep{2019ascl.soft05007B} to obtain a cutout of $31\times31$ pixels from the calibrated full-frame images centered on TOI-5688 A. The light curve was generated using a `normal' aperture in \texttt{eleanor}, which tests various aperture sizes, determined by the magnitude of the target star, and adopts the aperture that minimizes the combined differential photometric precision. {The aperture used in \texttt{eleanor} is shown in \autoref{fig:tirspec}.} The details of observations with TESS are given in \autoref{tab:tess_sect}.

\begin{figure*}
    \centering
    \includegraphics[width=0.81\linewidth]{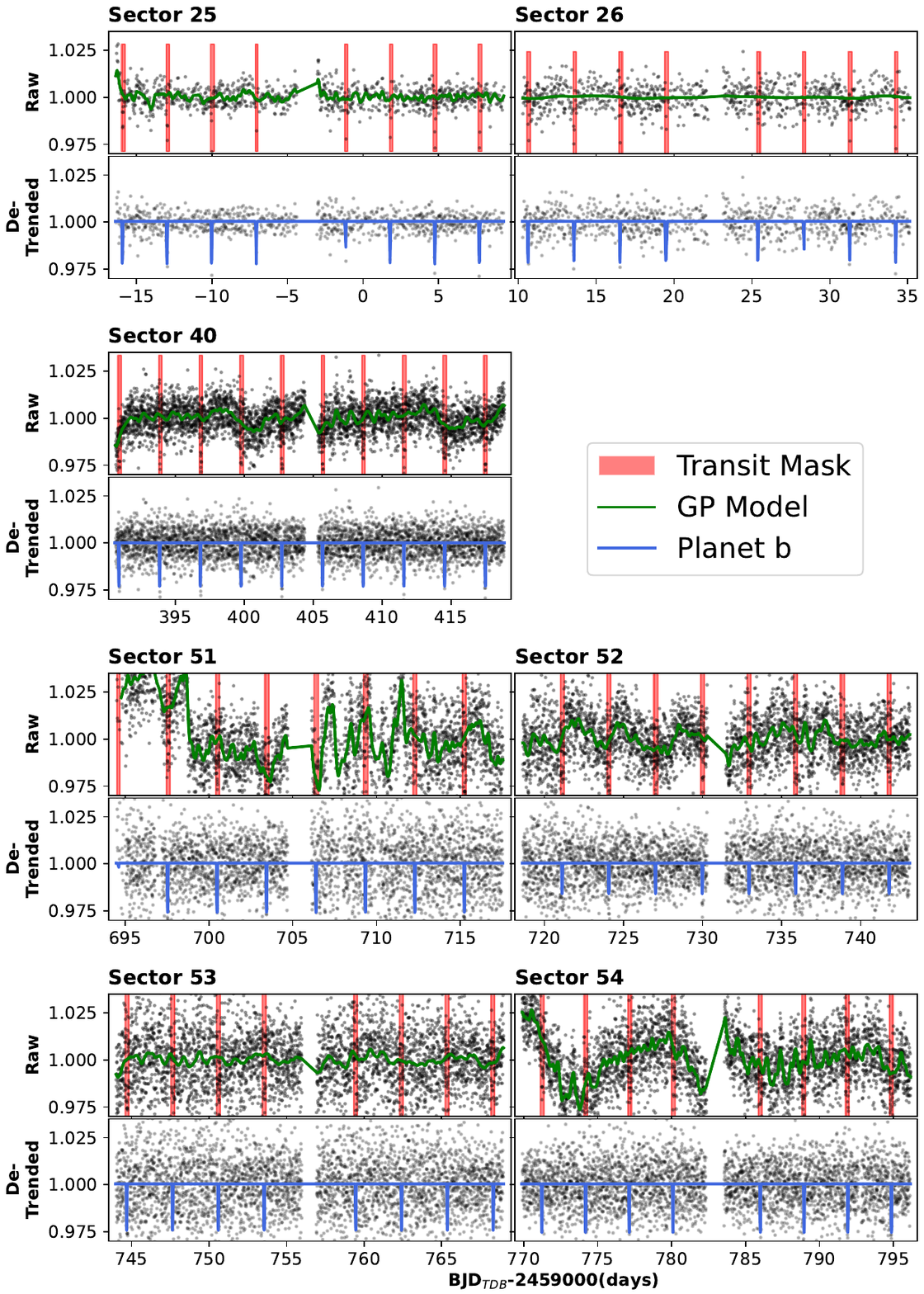}
    \caption{TESS \texttt{eleanor} light curves for sectors 25, 26, 40, 51, 52, 53 and 54. The first two sectors (25 and 26) have an exposure time of 30 minutes, and the rest have 10 minutes. The top panel in each figure shows the raw photometry data along with pink masks showing the transit regions. The mask is defined as $t_0\pm0.25$ days with $t_0$ as the transit center. The masked data was fit using a GP model, which was then subtracted (lower panel) before fitting the transit model.}
    \label{fig:raw_lc}
\end{figure*}

TESS's large pixels (21"/pixel) can often lead to source confusion, with multiple stars on the same pixel causing dilution. In addition, the long exposure times (30 min and 10 min) preclude accurate determination of the transit shape. To obtain more precise estimates and confirm the stellar host, we {obtained} ground-based transit observations {of TOI-5668~A~b}.

\subsection{Ground-based photometric follow up}
\label{sec:photometry}
\subsubsection{0.6 m Red Buttes Observatory}
\label{sec:rbo}
We observed TOI-5688 A b with the 0.6 m f/8.43 Ritchen-Chrétien Cassegrain at Red Buttes Observatory \citep[RBO;][]{Kasper_2016} in Wyoming, USA on 2022 October 09 and 2023 May 15. We used the  AltaF16 camera with a gain of 1.39 e\textsuperscript{-}/ADU, a plate scale of 0.731"/pixel, and $2\times2$ pixel on-chip binning. The target airmass ranged from 1.06 -- 2.3 on 2022 October 9, and 1.01 -- 1.31 on 2023 May 15.  The light curve was extracted from the frames of both data sets using a modified version of the pipeline outlined in \cite{2017AJ....153...96M}.{We included scintillation noise in quadrature to the photometric uncertainty}, as explained in \cite{Stefansson_2017}. The final extraction was done with an aperture radius of 3 pixels (2.19"), inner sky radius of 20 pixels (14.6"), and outer sky radius of 40 pixels (29.2"). The RBO observation details are given in \autoref{tab:tess_sect} and the light curves are shown in \autoref{fig:phase_folded}. 

\subsubsection{1.0 m Table Mountain Facility of Pomona College}
\label{sec:pom}
We also used the 1.0 m telescope of Pomona College residing at NASA JPL’s Table Mountain Facility (TMF), Wrightwood, California, USA \citep{2004AAS...20516804P} for photometric observations of TOI-5688 A on 2023 July 19. The airmass ranged between 1.03 and 1.07, with the observations being limited by twilight. The observations were carried out under $1\times1$ binning, a gain of 0.8 e\textsuperscript{-}/ADU, and a plate scale of 0.232"/pixel.

\begin{figure*}[t]
    \centering
    \includegraphics[width=0.9\linewidth]{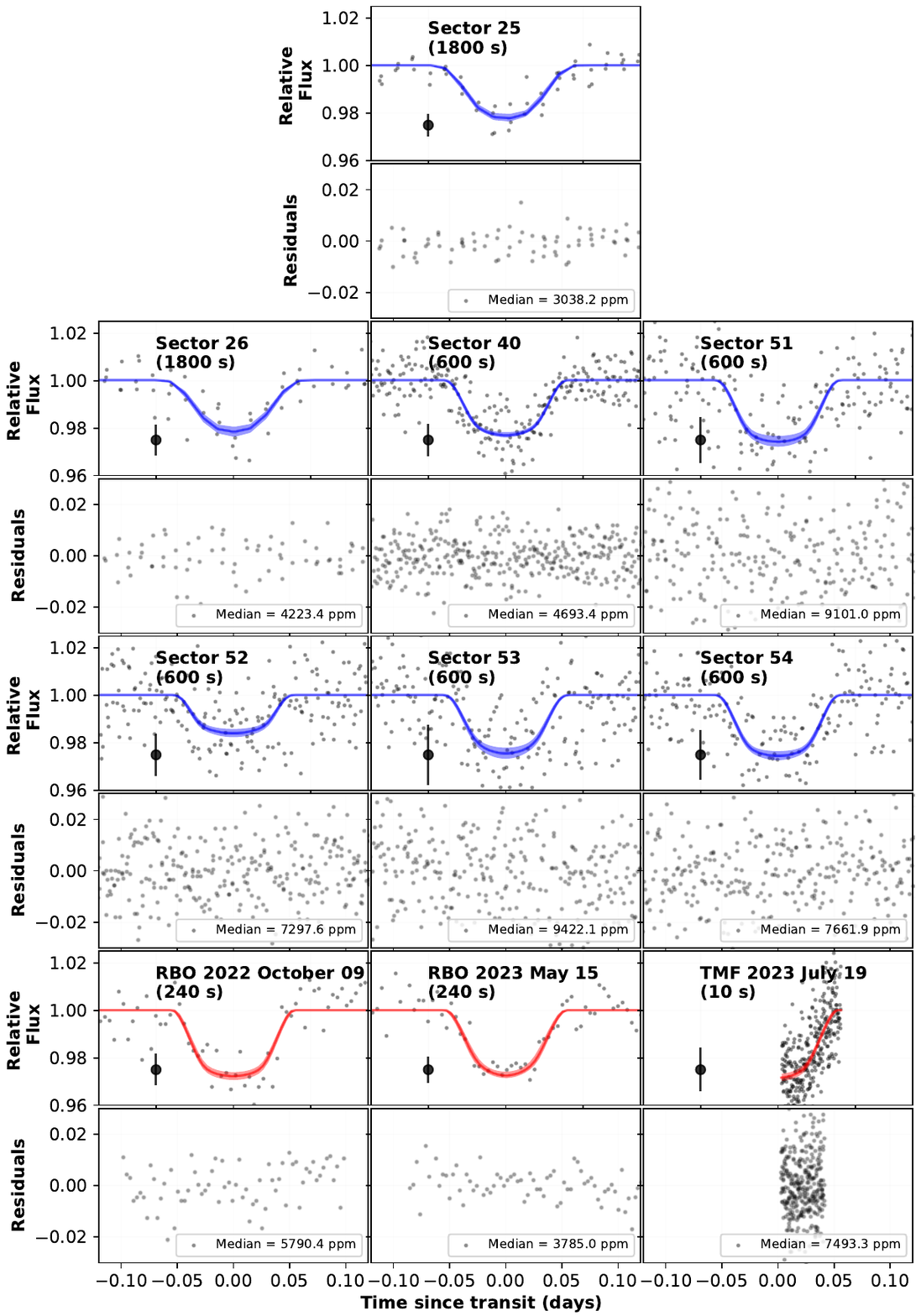}
    \caption{Phase folded transit light curves for TOI-5688 A b. Grey points represent the raw data. The model is shown in blue color for TESS and in red color for ground-based observations, along with the $1\sigma$ confidence intervals as translucent bands. The median statistical uncertainty is also shown at $x=-0.05$.}
    \label{fig:phase_folded}
    \vspace{-60pt}
\end{figure*}

The light curve from this visit was extracted using \texttt{AstroImageJ} \citep{astroimagej}, with an aperture radius of 15 pixels (3.39"), inner sky radius of 25 pixels (5.65"), and an outer sky radius of 30 pixels (6.78"). The average FWHM of PSF of TOI-5688 A in this data is $\sim 4.6"$. The observation parameters are listed in \autoref{tab:tess_sect}, and the light curve is shown in \autoref{fig:phase_folded}.

\subsubsection{uTIRSPEC at the {2.0} m Himalayan Chandra Telescope}
\label{tirspec}
\begin{figure}
    \centering
    \includegraphics[width=\linewidth]{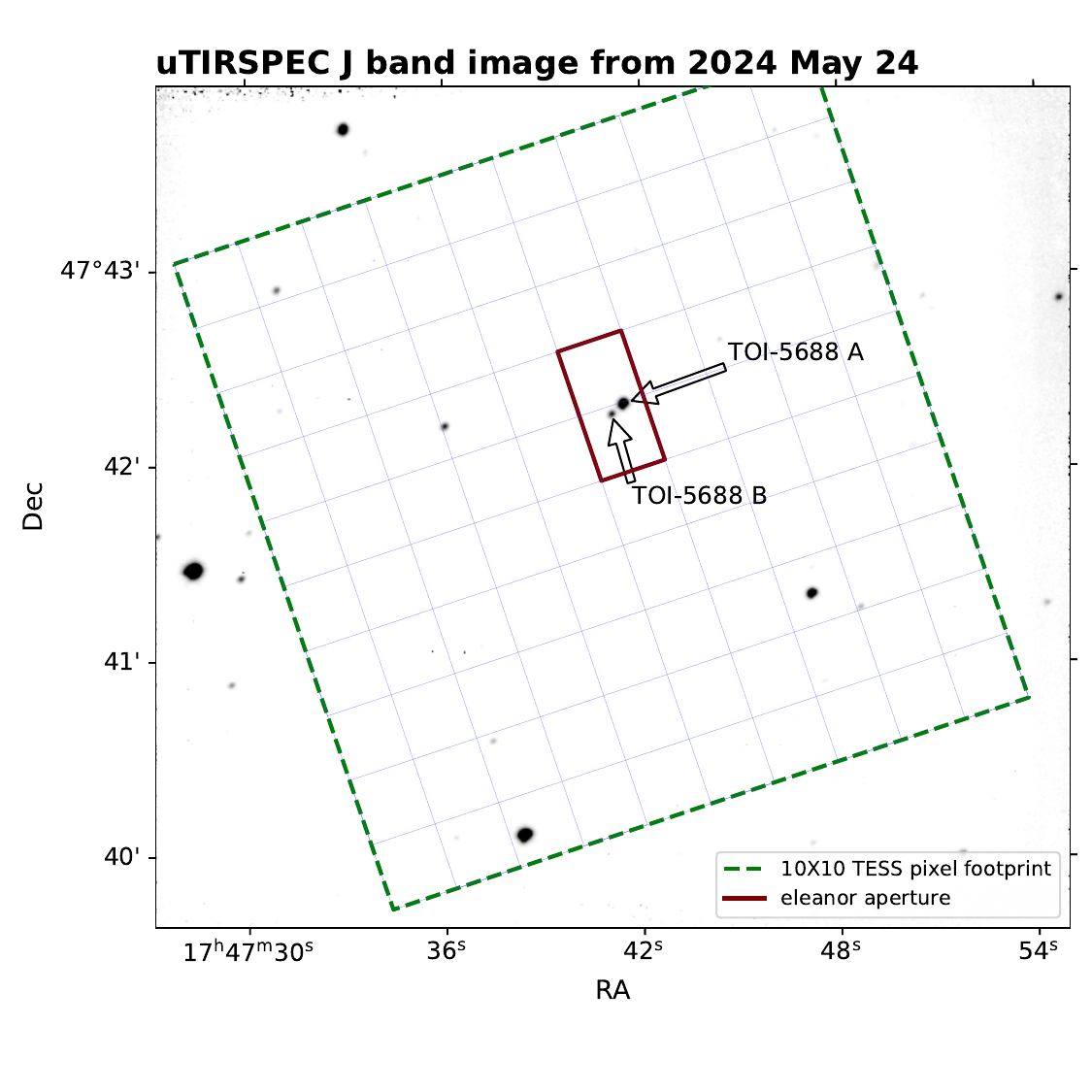}
    \caption{{J band image of TOI-5688 A and B taken with uTIRSPEC. A $10\times10$ pixel footprint from TESS Sector 40 is overlaid with the uTIRSPEC image, represented by the blue grid. The TESS aperture is highlighted in red, and both TOI-5688 A and B are marked. TESS observations of TOI-5688 A are contaminated by the presence of TOI-5688 B, making ground-based observations necessary to determine the true transit depth accurately.}}
    \label{fig:tirspec}
\end{figure}

We used the upgraded TIFR Infra-Red Spectrograph and Imager (uTIRSPEC) in its imaging mode to obtain near-infrared photometry of the TOI-5688 system.
The instrument is mounted on the 2 m Himalayan Chandra telescope, Hanle, Ladakh, India. Recently, the instrument TIRSPEC \citep{ninan2014tirspectifrnear} was upgraded to uTIRSPEC by replacing the HAWAII-1 PACE array with an H1RG array. The field of view in the imaging mode of uTIRSPEC is $5\times5$ arcmin\textsuperscript{2}. During the commissioning of uTIRSPEC, on 2024 May 24, we observed the TOI-5688 region. Multiple frames of 10-second exposures were taken in 2MASS J, H, and K\textsubscript{s} filters at 5 dither positions for good sky subtraction. Data were processed using the package \texttt{HxRGproc} \citep{hxrgproc}, after upgrading it to support frames taken with uTIRSPEC. After sky subtraction and flat correction, the images taken from multiple dither positions were shifted and combined. The function \texttt{DAOStarFinder} and  DAOPHOT algorithm \citep{1987PASP...99..191S}, in \texttt{photutils} \citep{larry_bradley_2024_10967176} were used to identify the sources in the frame. PSF photometry of the stars in the field was done using the function \texttt{PSFPhotometry} \citep{larry_bradley_2024_10967176} with an effective PSF \citep{2000PASP..112.1360A} of an aperture {with FWHM} $\sim6$ pixels ($\sim 1.8"$). {The PSFs are slightly elongated due to windy conditions, and therefore the effective PSF was derived using bright field stars and the \texttt{EPSFBuilder} function.}The instrument magnitudes of each source in J, H, and K\textsubscript{s} bands were calculated{ using PSF photometry.} The conversion from instrument magnitudes to apparent magnitudes (with color correction) {was subsequently performed }by cross-calibrating with the field stars 2MASS J, H, and K\textsubscript{s} magnitude. While the magnitudes we estimated for TOI-5688 A and B are consistent with 2MASS magnitudes, due to better spatial resolution, uTIRSPEC magnitudes are not affected by blending,{ specifically for TOI-5688 B.} The J, H, and K\textsubscript{s} magnitudes estimated with uTIRSPEC are given in \autoref{tab:stellarparamA}, with a J-band image of TOI-5688 A and B taken with uTIRSPEC {with TESS pixel footprint is} shown in \autoref{fig:tirspec}.

\subsection{Radial velocity observation with HPF and NEID}\label{sec:hpfrvs}
\subsubsection{HPF}
\label{sec:hpf}
{We obtained NIR spectra for TOI-5688~A }using the Habitable-zone Planet Finder \citep[HPF;][]{Mahadevan_2012, Mahadevan_2014, 10.1117/12.2313835} spectrograph {to measure its RVs}. The instrument is a fiber-fed \citep{fiberfed} near-infrared ($8080-12780$ \AA) high resolution ($R \sim 55,000$) precision RV spectrograph with a stabilized environment \citep{2016ApJ...833..175S}, {and }is mounted on the 10~m Hobby-Eberly telescope \citep[HET;][]{1998SPIE.3352...34R, 2021AJ....162..298H} at McDonald Observatory, Texas, USA. The telescope is a fixed-altitude range telescope with a roving pupil design, and it is fully queue-scheduled. The observations were carried out by HET resident astronomers \citep{2007PASP..119..556S} over 18 nights between 2022 August and 2023 July. Two 969 exposures were taken during each visit. {We performed the }bias correction, nonlinearity correction, cosmic ray reduction, and calculation of slope image and variance for each raw data frame separately, using the algorithms described in the package \texttt{HxRGproc} \citep{hxrgproc}. We use \texttt{barycorrpy} \citep{barycorrpy} to perform the barycentric correction on the individual spectra, which is the Python implementation of the algorithms from \cite{2014PASP..126..838W}. {We did not perform simultaneous NIR Laser Frequency Comb (LFC) calibrations } \citep{2019Optic...6..233M} due to concerns about the impact of scattered calibration light on our faint target. The wavelength solution for the target exposures was drift-corrected using the LFC exposures taken throughout the night of the observations. This approach has been demonstrated to enable precise wavelength calibration and drift correction with a precision of $\sim 30$ cm s\textsuperscript{-1} per observation \citep{Stefansson_2020AJ}. This value is much smaller than our expected per-observation RV uncertainty (instrumental + photon noise) for this object ($\sim45.7$ m s\textsuperscript{-1}). After binning for one night, this uncertainty is $\sim31.2$ m s\textsuperscript{-1}.

\begin{figure*}[t]
    \centering
    \subfigure[RV points]{
    \includegraphics[width=0.55\linewidth]{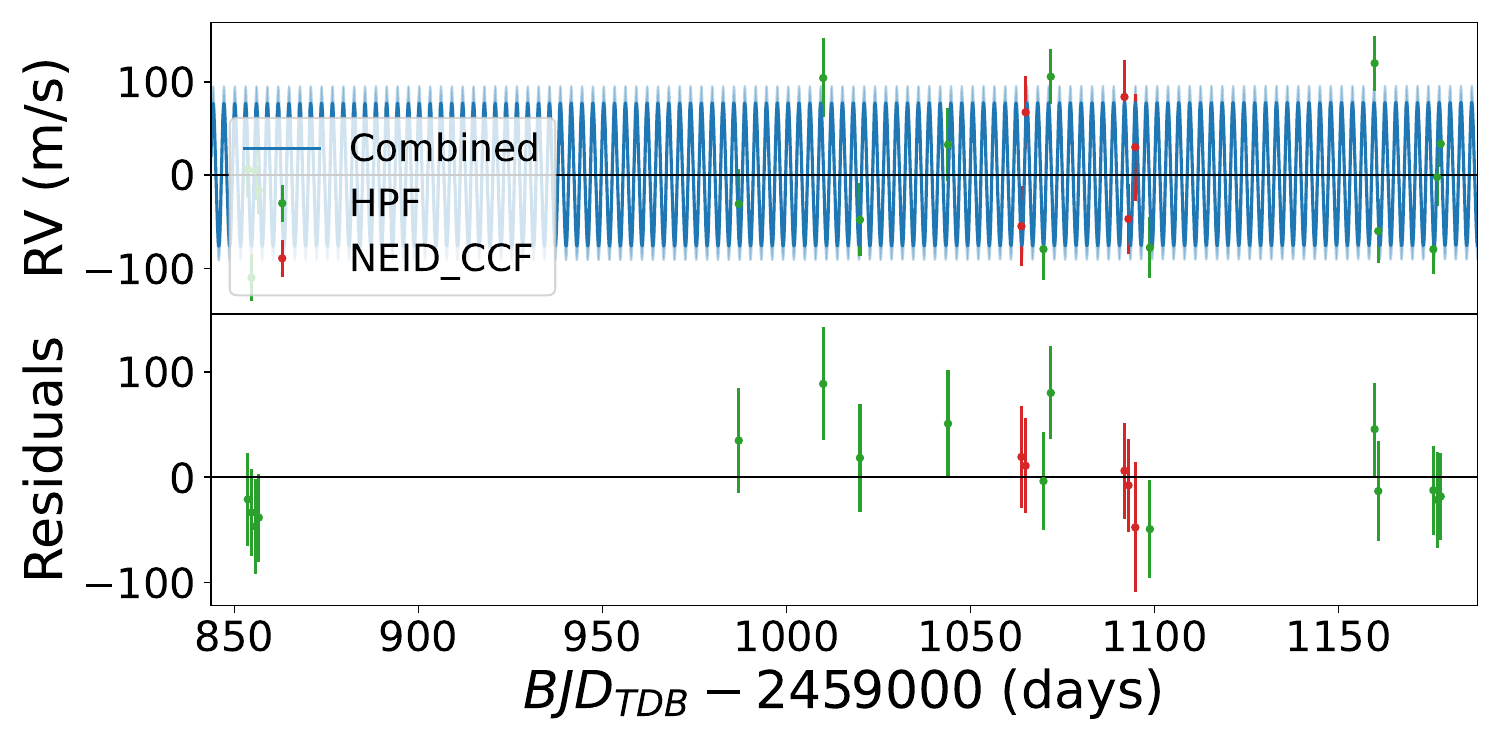}
    \label{fig:rv-all}
    }
    \subfigure[RV Phase folded]{\includegraphics[width=0.4\linewidth]{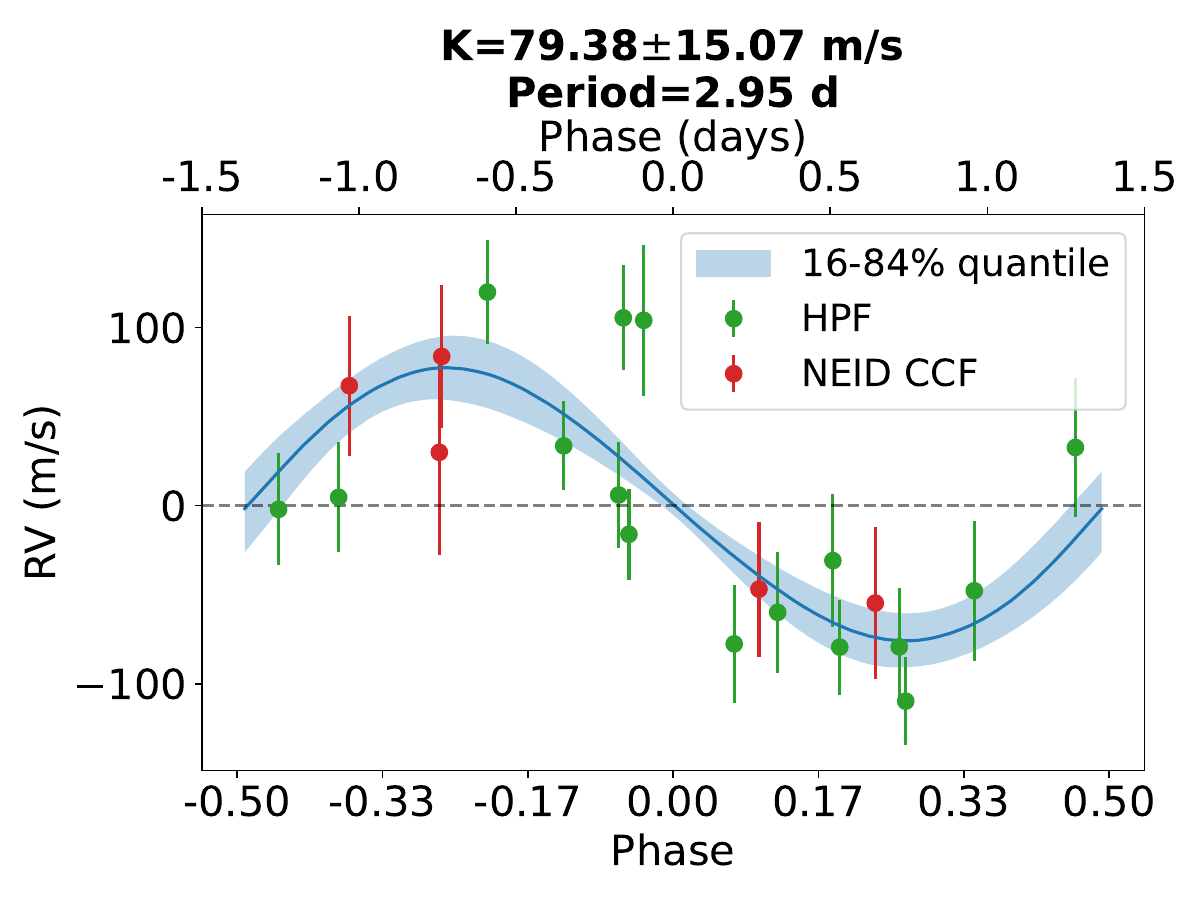}
    \label{fig:rv-phase}}
    \caption{\label{fig:rv-both} (a) Series of RV measurements of TOI-5688 A with HPF (green) and NEID (red). The values are given in \autoref{tab:org17b1d8f}. The best-fitting model from the joint fit of the transit and RVs is plotted in blue, including a 16\%--84\% confidence interval in light blue. The bottom panel shows the residuals after subtracting the model. (b) Phase-folded RV observations at the best-fit orbital period from the joint fit from Section \ref{sec:joint}. The eccentricity was not constrained in the fit, however, the results are consistent with a circular orbit (\autoref{tab:orbitalparam}).}
    
\end{figure*}

\subsubsection{NEID}
\label{sec:neid}
We also observed TOI-5688 A using NEID \citep{Schwab_2016,neid-main}, an ultra-precise environmentally stabilized \citep{neid-stable} spectrograph at the WIYN 3.5 m telescope\footnote{The WIYN Observatory is a joint facility of the NSF's National Optical-Infrared Astronomy Research Laboratory, Indiana University, the University of Wisconsin-Madison, Pennsylvania State University, Purdue University and Princeton University.} at Kitt Peak National Observatory in Arizona, USA. NEID has a fiber-fed system similar to HPF \citep{neidfibrefed}, with three fibers - science, sky, and calibration. The instrument has an extended red-wavelength coverage \citep[380-930 nm;][]{neid-design}. We observed the system for five nights between 2023 April 29 and 2023 May 31 using the high-resolution mode of NEID, with a resolution $R\sim 110,000$. In our analysis, we used the spectra with an exposure time of 1800 s with a median SNR per 1D extracted pixel of 5.2 at 850 nm. The median uncertainty in RV values is 39.9 m s\textsuperscript{-1}. The NEID observations on this target were part of a pilot program to test the faintness limit of the instrument for early M-dwarfs. 

The NEID data were reduced using the NEID data reduction pipeline\footnote{\url{https://neid.ipac.caltech.edu/docs/NEID-DRP/overview.html}} (DRP), and the level-2 1D extracted spectra were retrieved from the NEID archive\footnote{\url{https://neid.ipac.caltech.edu/search.php}}. The RVs were calculated using Cross-Correlation Functions with line mask\footnote{\url{https://neid.ipac.caltech.edu/docs/NEID-DRP/algorithms.html\#cross-correlation-based-rvs}}. Since the spectra are photon-noise limited and the number of observations is low, we {do not utilize the \texttt{SERVAL} template matching algorithm, which requires a high S/N template.}   The RVs obtained from HPF and NEID that are used for the analysis are shown in \autoref{tab:org17b1d8f}, with the phase folded model shown in \autoref{fig:rv-both}.


\begin{table}[htbp]
\caption{\label{tab:org17b1d8f}RV estimates of TOI-5688 A, taken with HPF and NEID. The RV values from HPF are binned down to one day.}
\centering
\begin{tabular}{cccc}
\hline
\hline
BJD\textsubscript{TDB}(days) & RV(m s\textsuperscript{-1}) & \(\sigma\) (m s\textsuperscript{-1}) & Instrument\\[0pt]
\hline
2459853.62080 & 60.2 & 29.8 & HPF \\[0pt]
2459854.60840 & -54.4 & 24.8 & HPF \\[0pt]
2459855.60560 & 60.6 & 31.0 & HPF \\[0pt]
2459856.60440 & 39.2 & 25.7 & HPF \\[0pt]
2459987.02480 & 23.3 & 37.3 & HPF \\[0pt]
2460009.95930 & 159.0 & 42.5 & HPF \\[0pt]
2460019.94120 & 3.8 & 39.3 & HPF \\[0pt]
2460043.87410 & 84.8 & 39.3 & HPF \\[0pt]
2460069.80140 & -20.6 & 33.1& HPF \\[0pt]
2460071.80010 & 163.7 & 29.5 & HPF \\[0pt]
2460098.71520 & -29.3 & 32.9& HPF \\[0pt]
2460159.77750 & 177.8 & 29.4& HPF \\[0pt]
2460160.77600 & -5.8 & 34.2& HPF \\[0pt]
2460175.73040 & -17.5 & 26.7& HPF \\[0pt]
2460176.74800 & 53.3 & 31.5& HPF \\[0pt]
2460177.72890 & 95.9 & 25.0& HPF \\[0pt]
\hline
2460063.82277 & -70.6 & 42.9 & NEID\\[0pt]
2460064.96150 & 51.4 & 39.3 & NEID\\[0pt]
2460091.81267 & 67.9 & 39.9 & NEID\\[0pt]
2460092.90407 & -62.8 & 37.9 & NEID\\[0pt]
2460094.75248 & 14.1 & 57.5 & NEID\\[0pt]
\hline
\end{tabular}
\end{table}

\subsection{Speckle Imaging with NESSI at WIYN}
\begin{figure}
    \centering
    \includegraphics[width=1.1\linewidth]{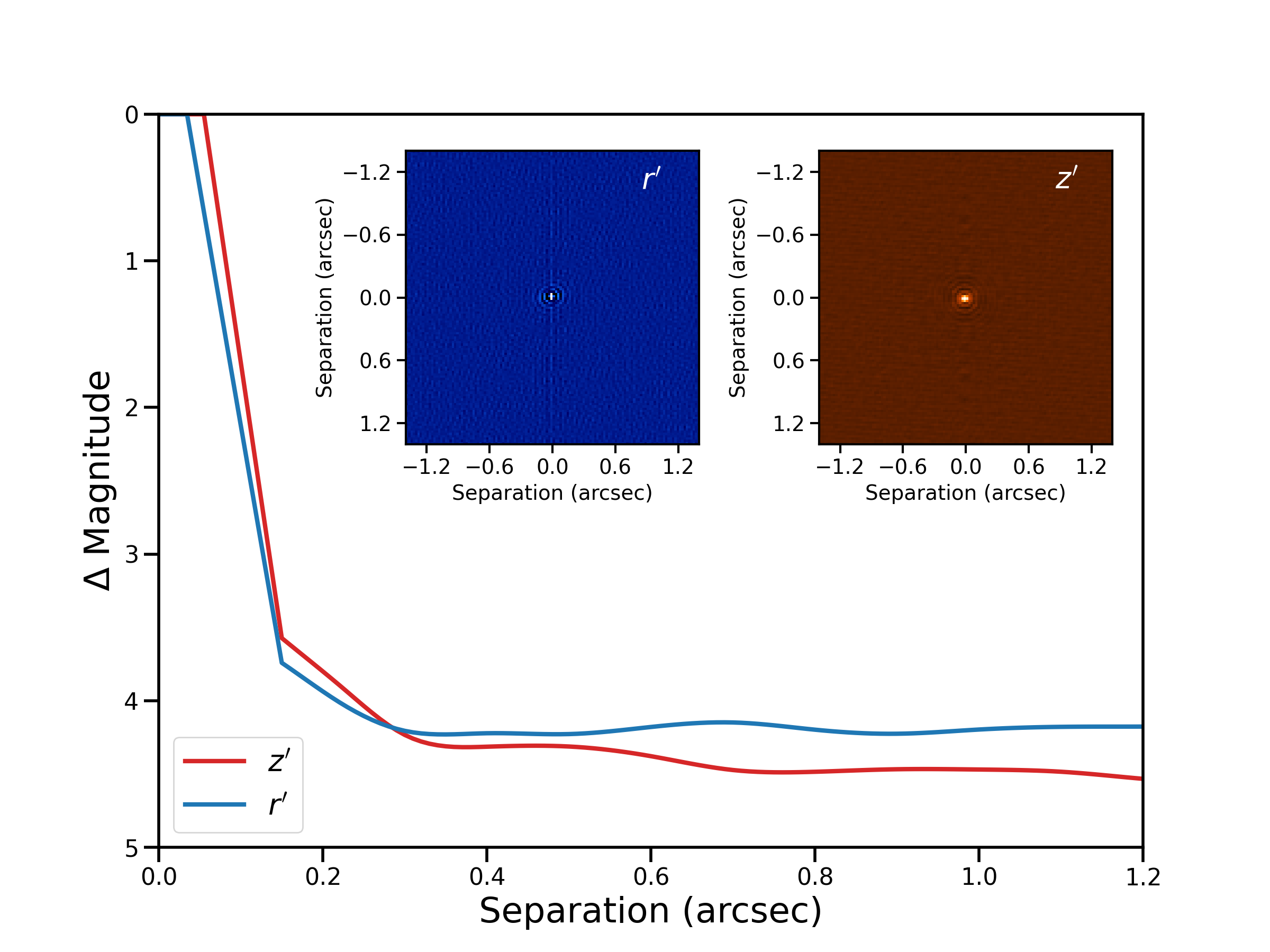}
    \caption{NESSI Speckle Imaging in r' and z' bands in the inset 2.4" across. The curve shows the 5-$\sigma$ contrast curve for TOI-5688 A in both $z$ and $r$ bands. The contrast curves indicate no bright companions within $1.2"$ from the host star.}
    \label{fig:specle}
 \end{figure}

We conducted observations of TOI-5688 A on 2022 September 16 with the NN-Explore Exoplanet Stellar Speckle Imager \citep[NESSI;][]{nessi} mounted on the WIYN 3.5 m telescope at Kitt Peak National Observatory, to identify faint background stars and nearby stellar companions. We acquired a sequence of 40 ms diffraction-limited exposures spanning 9 minutes, employing the SDSS \textit{r'} and SDSS \textit{z'} filters on NESSI. The speckle images were reconstructed using the methods outlined in \cite{speckleimg}. We did not detect any stellar sources {fainter than} $\Delta r' = 4.0$ or $\Delta z' = 4.0$ at separations $<1.2"$, as illustrated in \autoref{fig:specle}.
 
\section{Stellar Parameters}\label{sec:stellar}
\subsection{Wide Separation Binary and Galactic Kinematics}
\label{subsec:galactic}
The TOI-5688 system consists of a wide-separation binary, in which the stars are $\sim 1110$ AU \citep[$5.1"$;][]{riello2021} apart in the projected sky plane. {The values of proper motion and parallax of both TOI-5688 A and B are consistent (\autoref{tab:stellarparamA}) suggesting that they are co-moving stars. The catalog from \cite{el-bardy2021} also lists this system as binary. Using the values given in the Gaia DR3 catalog, the galactic velocities U, V, and W values of TOI-5688 A are calculated with the \texttt{Python} package \texttt{galpy}\footnote{\url{http://github.com/jobovy/galpy}} \citep{galpy_2015}. These values are consistent with the system being part of the galaxy's thin disk.}

{We used the parameter \texttt{phot\_bp\_rp\_excess\_factor} to determine if any {Gaia $B_P$ or $R_P$ are} contaminated by the companion. Using Equation 6 and Table 2 from \cite{riello2021}, we calculated the corrected \texttt{phot\_bp\_rp\_excess\_factor}, accounting for this parameter's color-dependent mean trend. The corrected \texttt{phot\_bp\_rp\_excess\_factor} for TOI-5688 B is $\approx0$, implying that the contamination by the companion is negligible in Gaia photometry.}

The companion star (TOI-5688 B) does not contaminate the spectroscopic observations from HPF and NEID with the on-sky fiber sizes are $1.7''$ \citep{fiberfed} and $0.9"$ \citep{neidfibrefed}, respectively. Our ground-based observations confirmed that the planet is orbiting the brighter and more massive primary, TOI-5688 A. The stellar parameters for TOI-5688 A and B are listed in \autoref{tab:stellarparamA}.

\subsection{Stellar parameters of TOI-5688 A}
\label{sec:mainstar}

We estimate the stellar parameters for TOI-5688 A using \texttt{HPF-SpecMatch}, available broadband photometry, and Gaia astrometry.  \texttt{HPF-SpecMatch} \citep{2020AJ....159..100S} is a \texttt{Python} package used to determine empirical stellar parameters from HPF spectra, using the template-matching algorithm described in \cite{specmatch}.
The spectral matching was performed on HPF order index 5 (8534 -- 8645 \AA) spectra that have minimal atmospheric contamination. TOI-5688 A is determined to have T\textsubscript{eff} $ = 3713\pm59$ K, [\text{Fe/H}]$ = 0.47\pm0.16$ dex and \logg $ = 4.69\pm0.04$. The resolution limit of HPF ($R\sim 55,000$) places a constraint of $v \sin i < 2$ km s\textsuperscript{-1}. The estimated stellar parameters are listed in \autoref{tab:stellarparamA}. 

{Even though \texttt{HPF-SpecMatch} provides a nominal metallicity estimate, we note the caveat that the \texttt{HPF-SpecMatch} template matching method estimates the stellar parameters by $\chi^2$ minimization of the entire order. This method is more sensitive to T\textsubscript{eff} and \logg, while not providing robust [Fe/H] estimates (essentially the $\chi^2$ valley is not narrow) for TOI-5688 A. Therefore, while we estimate metallicity of $0.47\pm0.16$ dex, we advise caution in interpreting this beyond being a super-solar metallicity M-dwarf.}


The mass and radius of the star TOI-5688 A were calculated by fitting the Spectral Energy Distribution (SED) with the package \texttt{EXOFASTv2} \citep{eastman2019exofastv2}. Due to the caution of blending with the companion and potential contamination in ground-based surveys because of the faintness of the target, only the Gaia \citep{gaiamag} magnitudes and 2MASS magnitudes estimated from the observations with uTIRSPEC (Section \ref{tirspec}) were used for SED fitting.

\subsection{Stellar parameters of TOI-5688 B}
\label{sec:companion}
We derive $\text{T\textsubscript{eff}} = 3294^{+91}_{-71}$ K, mass of $0.31\pm0.03$ M\textsubscript{$\odot$} and radius of $0.32\pm0.02$ R\textsubscript{$\odot$} for TOI-5688 B by fitting the SED with the package \texttt{EXOFASTv2} \citep{eastman2019exofastv2}. 
As in the case of TOI-5688 A, we only use Gaia \citep{gaiamag} and uTIRSPEC magnitudes for the SED fitting due to potential contamination in the magnitudes.

\begin{deluxetable*}{p{1.5cm}p{5cm}x{4.2cm}x{3.3cm}c}
\tablecaption{Summary of stellar parameters for TOI-5688 A and B. \label{tab:stellarparamA}}
\tablehead{\colhead{~~~Parameter}&  \colhead{Description (Unit)}&
\colhead{TOI 5688 A}&\colhead{TOI 5688 B}&
\colhead{Reference}}
\startdata
TIC & \tess{} Input Catalog &  193634953& 193634951 & Stassun\\
2MASS & $\cdots$ &  {J17474153+4742171} & {J17474118+4742138} &2MASS\\
Gaia DR3  & $\cdots$ & {1363205856494897024} & {1363205856494896896}&Gaia DR3\\
\hline
\multicolumn{4}{l}{\hspace{-0.2cm} Equatorial Coordinates, Proper Motion and} \\
~$\alpha_{J2016}$ & Right Ascension (RA) &    {17:47:41.54} & {17:47:41.18} & Gaia DR3\\
~$\delta_{J2016}$  & Declination (Dec) & {+47:42:18.17} & {+47:42:14.90} & Gaia DR3\\
~$\mu_{\alpha}$  & Proper motion (RA, \unit{mas/yr}) &  $-0.91 \pm 0.03$ & $-0.69\pm0.09$ & Gaia DR3\\
 ~$\mu_{\delta}$  & Proper motion (Dec, \unit{mas/yr})  & $64.93\pm 0.04$ & $64.96\pm0.09$  & Gaia DR3\\
  ~$\varpi$ & Parallax (mas) & $4.403\pm0.029$ &$4.339\pm0.064$& Gaia DR3\\
~$d$ &Distance in pc& $225.719^{+1.292}_{-1.510}$ & $230.582^{+3.543}_{-3.560}$  & Bailer-Jones\\
\hline
\multicolumn{4}{l}{\hspace{-0.2cm} Optical and near-infrared magnitudes:}  \\
~~~ $TESS$ & TESS mag & $14.21\pm0.01$ &$16.39\pm0.01$& TESS\\
~~~ $G$ & Gaia G magnitude & $15.3060\pm0.0006$ &$17.5562\pm0.0013$ & Gaia DR3 \\
~~~ $G_{BP}$ & Gaia BP magnitude & $16.4598\pm0.0057$ & $19.2407\pm0.0365$ & Gaia DR3 \\
~~~ $G_{RP}$ & Gaia RP magnitude & $	14.2249\pm0.0023$ & $16.3198 ± 0.0051$ & Gaia DR3 \\
 ~~~ $J$ & $J$ mag & $12.904\pm0.033$ & $14.594\pm0.031$ & uTIRSPEC\\
 ~~~ $H$ & $H$ mag & $12.22\pm0.05$ &  $14.00\pm0.05$ & uTIRSPEC\\
~~~ $K_s$ & $K_s$ mag & $12.01\pm0.05$ & $13.77\pm0.05$ & uTIRSPEC\\
\multicolumn{5}{l}{\hspace{-0.2cm} Stellar Parameters:}\\
~~~$T_{\mathrm{e}}$ &  Effective temperature (\unit{K}) & $3713\pm59$ &$3231^{+65}_{-62}$ & This work\\
~~~$\mathrm{[Fe/H]}$ &  Metallicity (dex) & $0.47\pm0.16$ & $0.20^{+0.13}_{-0.16}$ & This work\\ 
~~~ \logg & Surface gravity (cgs units) & $4.69\pm0.04$ &$4.92\pm0.04$& This work\\
~~~$M_\star$ &  Mass (M$_{\odot}$) & $ 0.60\pm0.02$ & $0.31\pm0.04$ & This work\textsuperscript{a}\\
~~~$R_\star$ &  Radius (R$_{\odot}$) & $0.57\pm0.02$ &$0.32\pm0.02$ & This work\textsuperscript{a}\\
~~~$L_\star$ &  Luminosity (L$_{\odot}$) & $0.0591^{+0.0027}_{-0.0018}$ &$0.01082\pm0.0007$& This work\textsuperscript{a}\\
~~~$\rho_*$ &  Density (g/cm\textsuperscript{3}) & $4.43^{+0.31}_{-0.30}$ &$ 13.4^{+2.3}_{-1.7}$& This work\\
~~~Age & Age (Gyrs) & $7.2^{+2.8}_{-4.0}$ & $9.7^{+2.7}_{-4.1}$ & This work\\
\multicolumn{4}{l}{\hspace{-0.2cm} Galactic Parameters:}           \\
~~~$\Delta$RV &  ``Absolute'' radial velocity (\unit{km/s}) & $-83.3\pm0.1$ &$\cdots$& Gaia DR3\\
~~~$U, V, W$ &  Galactic velocities (\unit{km/s}) &  $-86.56\pm0.45$,  $-53.73\pm0.12$, $-30.53\pm0.08$ & $\cdots$ & This work\\ 
~~~$U, V, W$\textsuperscript{b} & LSR Galactic velocities (\unit{km/s}) & $-75.46\pm0.96$, $-41.49\pm0.70$, $-23.28\pm0.61$ & $\cdots$ & This work\\ 
\enddata
\tablenotetext{}{References: Stassun \citep{2018AJ....156..102S}, Gaia DR3 \citep{2018yCat.1345....0G, gaiamag},  Bailer-Jones \citep{vo:gedr3dist_main, 2023AJ....166..269B}, TIRSPEC \citep{ninan2014tirspectifrnear}} 
\tablenotetext{a}{\texttt{EXOFASTv2} derived values using MIST isochrones with the Gaia parallax as priors.}
\tablenotetext{b}{The barycentric UVW velocities are converted into local standard of rest (LSR) velocities using the constants from \cite{2010MNRAS.403.1829S}.}
\end{deluxetable*}

\section{Joint Fitting of Photometry and RVs}\label{sec:joint}
We jointly fit the {transit and RV} data using the \texttt{exoplanet} \citep{exoplanet:joss} package, which uses the NUTS sampling \citep[No U-Turn Sampling;][]{hoffman2011nouturn} {in the} Hamiltonian Monte Carlo \citep[HMC;][]{betancourt2018conceptual} method for posterior estimation with the \texttt{PyMC3} package \citep{10.7717/peerj-cs.55}. The \texttt{exoplanet} package models the transits with the package \texttt{starry} \citep{2018ascl.soft10005L}, which uses a quadratic limb-darkening law \citep{2002ApJ...580L.171M} parameterized for uninformative sampling, as explained in \cite{2013MNRAS.435.2152K}. We fit the transit in each TESS sector and ground-based observation using separate limb-darkening coefficients. {The TESS photometric fit includes a separate dilution factor \citep{Torres_2011} for each sector, which is constrained based on the ground-based observations (where the stars are spatially resolved). The fitted parameters are listed in \autoref{tab:orbitalparam}.} \\
To correct for the stellar and instrumental variability in the light curve, we mask out the transit signals and then fit a Gaussian Process (GP) model separately for each TESS sector. This model is then subtracted out (\autoref{fig:raw_lc}). We use the \texttt{RotationTerm} kernel \citep{celerite21}, which is implemented in \texttt{celerite2} \citep{celerite22} as the sum of two simple harmonic oscillators. {The standard deviation of the process ($\sigma$), the primary period of variability, the quality factor of the secondary oscillation ($Q_0$), the difference between the quality factors of primary and secondary modes ($dQ$), and the fractional amplitude of the secondary mode compared to the primary mode ($f$) are the hyperparameters of this model. The RV data were not used in this GP model fit of the TESS light curves.} The \texttt{exoplanet} package oversamples the time series during the model evaluation, to account for the long-cadence photometry of TESS.\\
We model the RVs with the standard Keplerian model with free eccentricity and omega that were sampled using a prior distribution \textit{Unit Circle}. We include an instrument RV offset and a linear trend for the entire RV time series.  We also include a simple white-noise jitter term in quadrature to measure the stellar RV and photometry jitter from each dataset. \\
{We use \texttt{scipy.optimize} \citep{scipy} to get }maximum a posteriori estimates as the initial condition for posterior sampling. Four chains consisting of 9000 steps (6000 tune + 3000 draw) in each chain were sampled for HMC.  The convergence of sampling was checked using the Gelman-Rubin Statistic \citep[$\hat{R}\leq1.1$;][]{2006ApJ...642..505F}. The estimated system parameters are listed in \autoref{tab:orbitalparam}.

\begin{deluxetable*}{lp{3cm}c}
\tablecaption{Summary of orbital and physical parameters for TOI-5688 A b. \label{tab:orbitalparam}}
\tablehead{\colhead{~~~Parameter}&  \colhead{Units}  & \colhead{Value\textsuperscript{a}}}
\startdata
\sidehead{Orbital Parameters:}
Orbital Period & P (days) & \(2.94815527^{+0.00000452}_{-0.00000448}\)\\[0pt]
Eccentricity & e & \(0.128^{+0.078}_{-0.077}\)\\[0pt]
Argument of Periastron & \(\omega\) (degrees) & \(-1.32^{+1.12}_{-0.86}\)\\[0pt]
Semi-amplitude Velocity & K (m/s) & \(79.4^{+14.6}_{-15.5}\)\\[0pt]
Systemic Velocity\textsuperscript{b} & \(\gamma_{\text{HPF}}, ~\gamma_{\text{NEID}}\) (m/s) & $54.2^{+12.5}_{-11.9},~ 30.6^{+25.0}_{-24.7}$\\[0pt]
RV trend & \(\frac{dv}{dt}\) (m/s) & $0.77^{+4.90}_{-5.94}$\\[0pt]
RV jitter & \(\sigma_{\text{HPF}}, ~\sigma_{\text{NEID}}\) (m/s) & $33.0^{+15.1}_{-13.3},~22.4^{+35.1}_{-16.0}$\\[0pt]
\hline
\sidehead{Transit Parameters:}
Transit Midpoint & T\textsubscript{C} (BJD\textsubscript{TDB}) & \(2459771.26024^{+0.00058}_{-0.00058}\)\\[0pt]
Impact parameter & b & \(0.714^{+0.046}_{-0.095}\) \\[0pt]
Scaled Radius & \(R_p/R_*\) & \(0.164^{+0.007}_{-0.009}\)\\[0pt]
Scaled Semimajor Axis & \(a/R_*\) & \(12.5^{+0.50}_{-0.45}\)\\[0pt]
Orbital Inclination & i (degrees) & \(87.05^{+0.31}_{-0.25}\)\\[0pt]
Transit Duration & T\textsubscript{14} (days) & \(0.0698^{+0.0044}_{-0.0036}\)\\[0pt]
Photometric Jitter\textsuperscript{c} & \(\sigma_{TESS,S25}\) (ppm) &\(242^{+270}_{-158}\)\\[0pt]
 & \(\sigma_{TESS,S26}\) (ppm) & \(403^{+448}_{-274}\)\\[0pt]
 & \(\sigma_{TESS,S40}\) (ppm) & \(2311^{+225}_{-233}\)\\[0pt]
 & \(\sigma_{TESS,S51}\) (ppm) & \(8620^{+266}_{-263}\)\\[0pt]
 & \(\sigma_{TESS,S52}\) (ppm) & \(5925^{+234}_{-228}\)\\[0pt]
 & \(\sigma_{TESS,S53}\) (ppm) & \(7105^{+326}_{-337}\)\\[0pt]
 & \(\sigma_{TESS,S54}\) (ppm) & \(4168^{+343}_{-355}\)\\[0pt]
 & \(\sigma_{RBO20221009}\) (ppm) & \(4497^{+1257}_{-1354}\)\\[0pt]
 & \(\sigma_{RBO20230515}\) (ppm) & \(875^{+1040}_{-610}\)\\[0pt]
 & \(\sigma_{TMF20230719}\) (ppm) & \(7881^{+728}_{-732}\)\\[0pt]
Dilution\textsuperscript{d} & \(D_{TESS,S25}\) & \(0.808^{+0.084}_{-0.075}\)\\[0pt]
 & \(D_{TESS,S26}\) & \(0.793^{+0.103}_{-0.093}\)\\[0pt]
 & \(D_{TESS,S40}\) & \(0.826^{+0.065}_{-0.062}\)\\[0pt]
 & \(D_{TESS,S51}\) & \(0.930^{+0.107}_{-0.096}\)\\[0pt]
 & \(D_{TESS,S52}\) & \(0.573^{+0.076}_{-0.068}\)\\[0pt]
 & \(D_{TESS,S53}\) & \(0.881^{+0.108}_{-0.099}\)\\[0pt]
 & \(D_{TESS,S54}\) & \(0.925^{+0.092}_{-0.085}\)\\[0pt]
 \hline
 \sidehead{Planetary Parameters:} 
 Mass & \(M_p\) (M\(_{\oplus}\) / M$_J$) & \(124.0^{+23.4}_{-24.4}~/~0.390^{+0.074}_{-0.077}\)\\[0pt]
Radius & \(R_p\) (R\(_{\oplus}\) / R$_J$) & \(10.3^{+0.6}_{-0.7}~/~0.920^{+0.053}_{-0.062}\)\\[0pt]
Density & \(\rho_p\) (g/cm\textsuperscript{-3}) & \(0.61^{+0.20}_{-0.15}\)\\[0pt]
Semimajor Axis & a (AU) & \(0.03379^{+0.00046}_{-0.00045}\)\\[0pt]
Average Incident Flux\textsuperscript{e} & $\langle F\rangle$ (10\textsuperscript{5}W/m\textsuperscript{2}) & \(0.66\pm0.02\)\\[0pt]
Planetary Insolation & S (S\(_{\oplus}\)) & \(50.3\pm5.0\)\\[0pt]
Equilibrium Temperature & $T_{eq}$ (K) & \(742\pm18\)\\
\enddata
\tablenotetext{a}{The reported value refer to the 16-50-84\% percentile of the posteriors.}
\tablenotetext{b}{In addition to the absolute RV given in \autoref{tab:org17b1d8f}}
\tablenotetext{c}{Jitter (per observation) added in quadrature to photometric instrument error.}
\textsuperscript{d}Dilution due to the presence of the background stars in the TESS aperture.\\
\tablenotetext{e}{We use the solar flux constant 1360.8 Wm\textsuperscript{-2} to convert insolation to incident flux.}

\vspace{-60pt}
\end{deluxetable*}

\section{Discussion}\label{sec:discussion}

\subsection{TOI-5688 A b in M dwarf planet parameter space}

\begin{figure}
    \centering
    \includegraphics[width=1.1\linewidth]{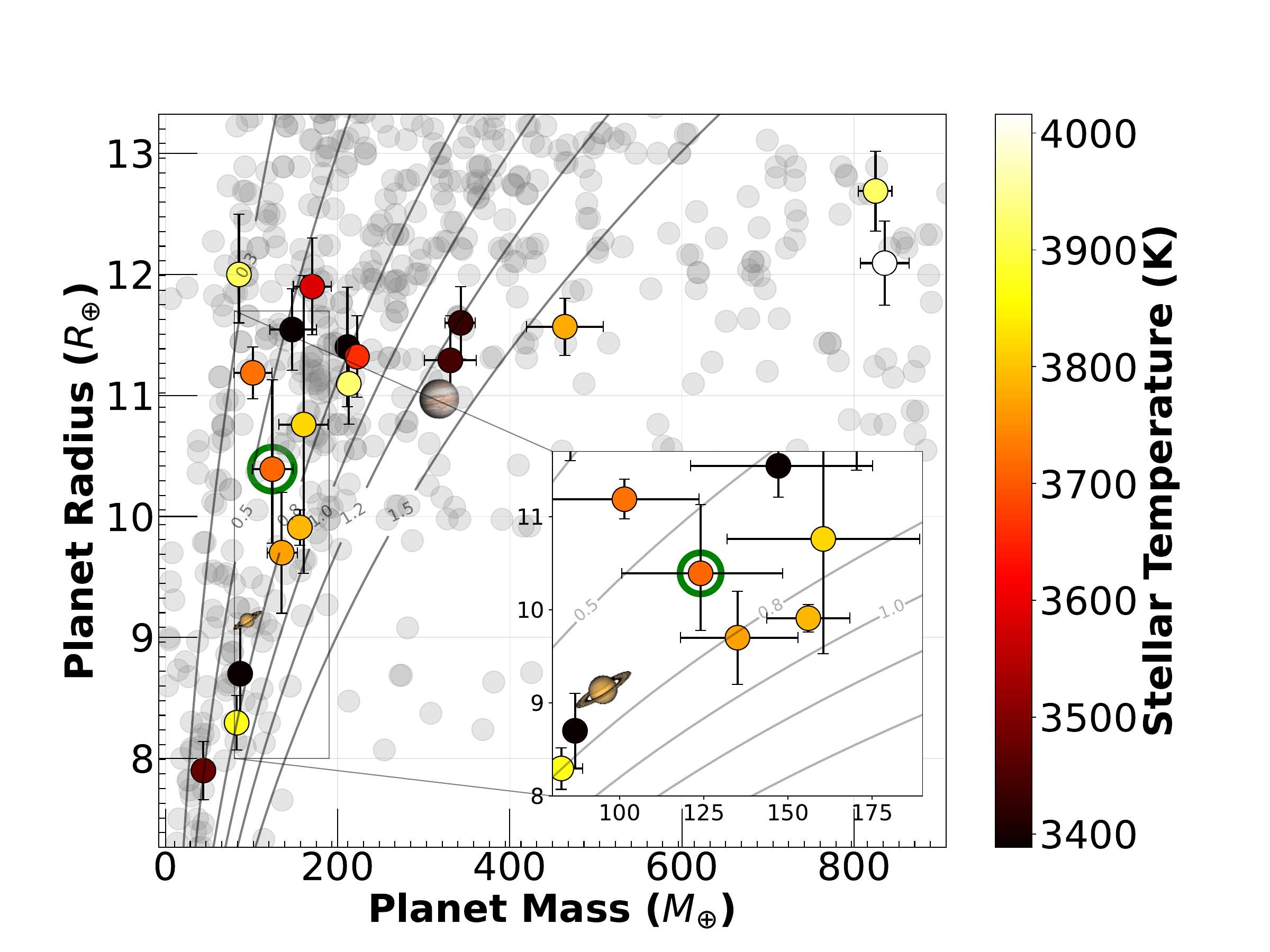}
    \caption{Parameter space of giant exoplanets hosted by M-dwarfs. TOI-5688 A b is marked by the green circle. Other M dwarf planets within the range of mass 95 -- 850 \earthmass{} are also shown in the figure, with colors representing the stellar effective temperature. The planets around FGK type stars are added in the background and represented by gray color along with the density contours for 0.3, 0.5, 0.8, 1.0, 1.2, and 1.5 g/cm\textsuperscript{3}.}
    \label{mass-radius}
\end{figure}

\begin{figure}[b]
    \centering
    \includegraphics[width=1.1\linewidth]{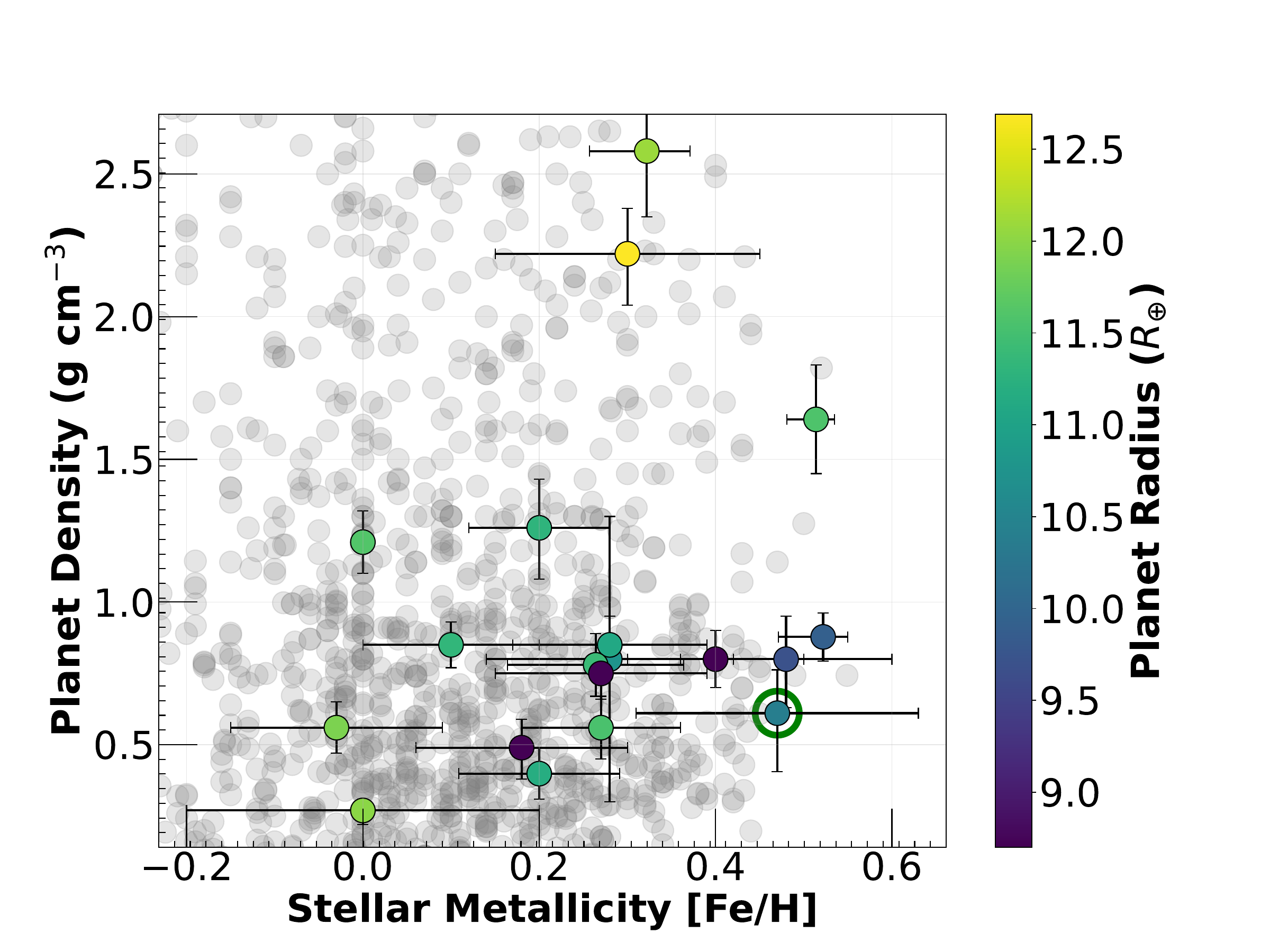}
\includegraphics[width=1.1\linewidth]{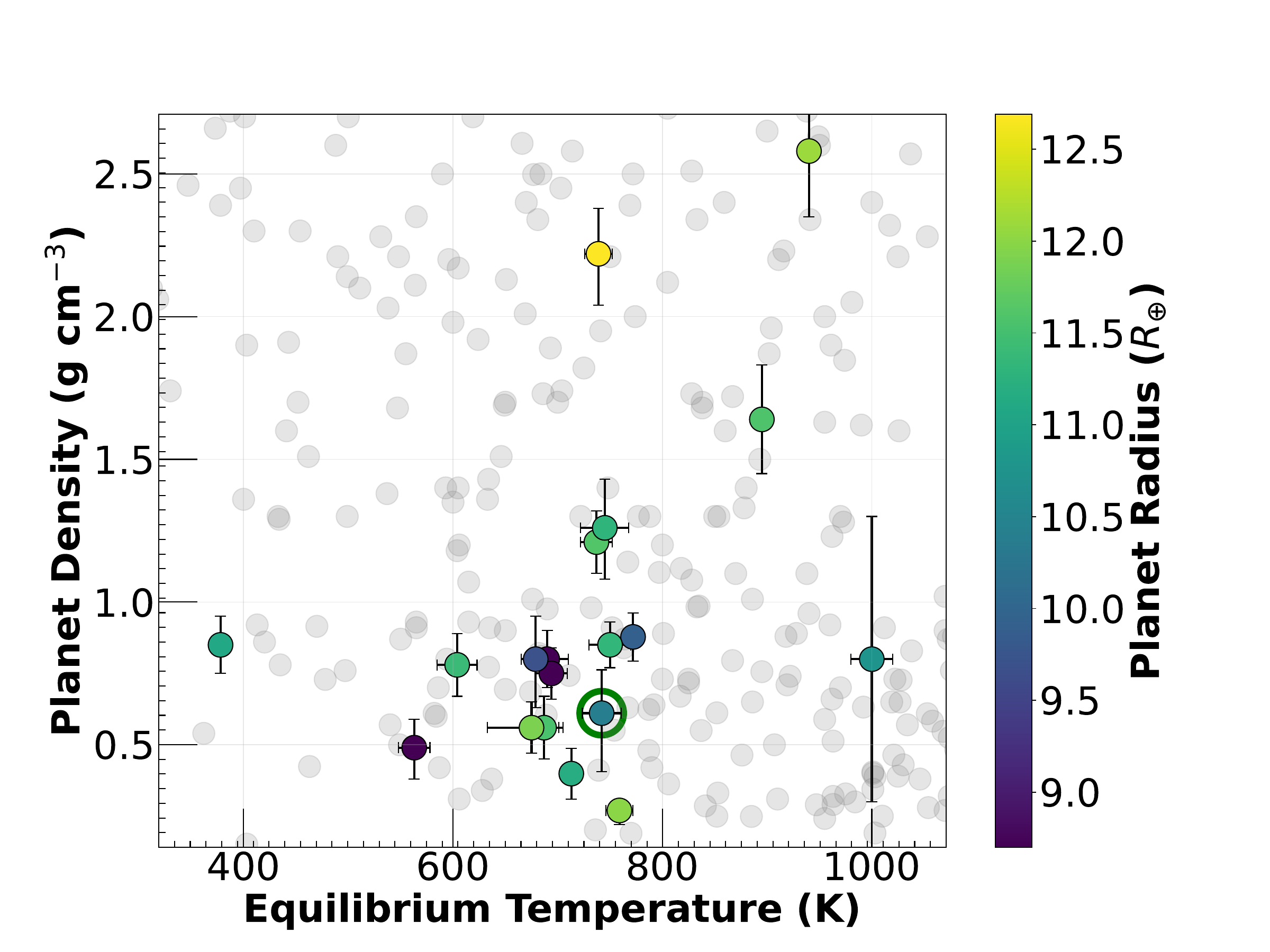}
    \caption{Plot of planet density with metallicity and equilibrium temperature of the planet. The upper plot shows the stellar metallicity-density space. The lower plot shows the equilibrium temperature-density space. The color of data points represents the radius of the planet. {Both the host-star metallicity and equilibrium temperature are likely to affect} the density of the planet.}
\label{fig:density}
\end{figure}

Around twenty-five giant exoplanets ($\gtrsim 8$ \earthradius) hosted by M-dwarfs ($\text{T\textsubscript{eff}} < 4000$ K) have been discovered so far. In this section, we compare the position of TOI-5688~A~b in the parameter space of transiting GEMS as queried from the NASA Exoplanet Archive \citep[NEA;][]{Akeson_2013} on 2024 May 20. The planets have $>3\sigma$ masses, and radii in the range of radius 8.0 -- 15.0 \earthradius. The host stars of the GEMS planets have an effective temperature in the range of 3300 -- 4000~K. Planets around FGK-type stars are shown in the background of the plots.

The mass vs. radius plot with density contours is shown in \autoref{mass-radius}. In addition, the positions of Saturn and Jupiter are shown. The data points are colored on the basis of the stellar effective temperature. TOI-5688~A~b has a mass of $124\pm24$~\earthmass~ and a radius of $10.4\pm0.7$ \earthradius, which is similar to that of Saturn with mass 95.2~\earthmass{} and radius 9.4~\earthradius{}. The planets HATS-6 b \citep{2015AJ....149..166H}, TOI-519 b \citep{2021AA...645A..16P}, Kepler-45 b \citep{2012AJ....143..111J}, TOI-5344 b \citep{Han_2023} and HATS 75 b \citep{Jordan_2022} are within 1$\sigma$ of the estimated mass of TOI-5688 A b. In addition, planets such as TOI-3629 b and TOI-4860 b are within the same density range as TOI-5688 A b and its neighbors. These planets have densities between 0.3 and 0.9 g/cm\textsuperscript{3}, masses in the range of 90 -- 150 \earthmass. Since the mass and density of Saturn in the solar system are in this range\footnote{\url{https://nssdc.gsfc.nasa.gov/planetary/factsheet/saturnfact.html}}, we collectively refer to these planets as `warm Saturn-like' planets. 

\subsection{Density of Saturn-like planets}
\label{subsec:dens_met}
The top panel of \autoref{fig:density} shows the metallicity of the host star and the density of the planet. The color of each data point represents the radius of the planet.

However, it is crucial to note the caveats { associated with the determination of M-dwarf metallicity. Given the forest of molecular lines in M-dwarf photospheres (due to their cooler temperatures) and the difficulty in continuum estimation, it is not possible to trivially use the standard metallicity estimation methods. Furthermore, \cite{2022A&A...658A.194P} show the perils of comparing M-dwarf metallicities obtained from different spectroscopic methods {(with different underlying assumptions)}. The metallicities included in \autoref{fig:density} are determined by a combination of photometric and spectroscopic methods, further complicating this. }Therefore, the discussions related to the metallicity of GEMS hosts should be interpreted with caution.

The bottom panel of \autoref{fig:density} shows the planetary equilibrium temperature (assuming 0 albedo) vs its bulk density. To check whether the equilibrium temperature of the planet drives the density, we calculated the Pearson correlation coefficient ($r=0.41$, p-value = 0.090) between the density of the planet and the equilibrium temperature in \autoref{fig:density}. This lack of statistically significant correlation suggests the absence of a direct correlation between these quantities for these cooler giant planets ($<$ 1000 K). 


\subsection{Formation Mechanism}
 Saturn-like exoplanets hosted by M-dwarfs are believed to have formed by the core accretion model of the planet formation \citep[etc]{core_accr2, POLLACK199662}, though their lower masses compared to Jupiters remains a mystery. One possible explanation by \cite{MOVSHOVITZ2010616} states that Saturn-like planets are formed by the slowing down of the runaway accretion from the disk as a result of the higher opacity of the disk. Recent studies of \cite{howard2023} and \cite{Helled_2023} also refer to Saturns as `failed giant planets'. \cite{Helled_2023} suggests that the `Saturns' took a few Myrs to form so that they had never gone through runaway accretion.   For super-solar metallicity stars, high metallicity can increase the disk's opacity. {This reduces the heat transfer efficiency \citep{2002ApJ...567L.149B}, which slows down gas accretion. This slowdown prevents planets from becoming excessively massive before the disk dissipates \citep{Helled_2023}. Consequently, it leads to the formation of planets in the bottom right portion of the top panel in \autoref{fig:density}.} Simulations from \cite{2006ApJ...643..501B, 2010ApJ...725L.145B} and most recently by \cite{2023ApJ...956....4B} have also explored the gravitational instability {pathway to form GEMS}. In the following subsections, we explore the potential formation pathways of TOI-5688 A b through both core accretion and gravitational instability. 

\subsubsection{Core accretion}
The formation of giant planets through core accretion is a two-step process. First, a rocky core of mass $\gtrsim 10$ \earthmass~ is formed by the coagulation of planetesimals \citep{POLLACK199662}, pebbles \citep{Lambrechts_2012} or both \citep{Alibert_2018}. Second, once the core is massive enough, i.e. $M_{\text{core}} \gtrsim 10$ \earthmass{} \citep{mordasini2007giant}, runaway gas accretion takes place onto the core from the protoplanetary disk, enabling the transition of a planet to a gas giant. We can separately analyze \textit{in-situ} and \textit{ex-situ} formation scenarios.
\paragraph{\textit{In-situ}}
The \textit{\textit{in-situ}} formation of the core with the required mass ($>10$ \earthmass) within the lifetime of the disk is challenging \citep{refId0, 2022ApJ...939L..10P,  2023ASPC..534..501M, 2024ApJ...963..122P}. \textit{In-situ} formation of the planet at its present-day orbital separation necessitates high disk surface density \citep[$\gtrsim7\times10^6$ g/cm\textsuperscript{2};][]{Dawson_2018} or a large feeding zone ($\Delta$, in terms of Hill radius; $>14000$). {Models predict a surface density of only $\sim 10^3$ g/cm\textsuperscript{2} at $r=0.03$ AU \citep{2023ASPC..534..501M}.} Secondly, $\Delta$ has a hard upper limit set by the escape velocity of the disk \citep{Schlichting_2014}, which is given by Equation 7 of \cite{Dawson_2018}. At {TOI-5688~A~b's present-day location at }P = 2.95 days, and assuming a core density of 4 g/cm\textsuperscript{3} \citep[the average density of Saturn's core; ][]{mankovich2021diffuse}, $\Delta_{max}$ is given by $$\Delta_{max}\simeq13\left(\frac{P}{3~\text{day}}\right)^{1/3}\left(\frac{\rho}{8~\text{g cm\textsuperscript{-2}}}\right)^{1/6} \simeq 11.5$$  Given the limitations in the size of the \textit{in-situ} feeding zone and disk surface density, it is highly improbable for TOI-5688 A b to have formed \textit{in-situ}.

\paragraph{\textit{Ex-situ}}
The planet could have formed farther out in the disk and subsequently migrated to its present location. We estimate the location that can support core formation using Equations 6 and 7 from \cite{Dawson_2018}. For the dust surface density profile, we adopt the power law $\Sigma_{dust}(r)=\Sigma_0\left(\frac{r}{\text{AU}}\right)^{-1}\exp\left(-\frac{r}{AU}\right)$ \citep{Williams_2011}, where $r$ denotes the distance from the star. 
$\Sigma_0$ is the normalization constant, which is the value of $\Sigma$ at 1 AU, and calculated using the relation $$\Sigma_0 = \frac{M_{dust}}{2\pi R_c^2}$$ Here, $R_c= 1$ AU. The disk mass ($M_{gas} + M_{dust}$) can be estimated to be $\sim 3996$ \earthmass, which is $2\% $ of the star \citep{flores2023earlyplanetformationembedded}. If the gas-to-dust ratio is 100 \citep{flores2023earlyplanetformationembedded}, $\Sigma_0$ can be estimated as $\sim 167$ g/cm\textsuperscript{2}. 
We assume the size of the maximum feeding zone, i.e., $\Delta=\Delta_{max}$, and substitute Equation 7 of \cite{Dawson_2018} and the power law of dust surface density into Equation 6 of \cite{Dawson_2018} to obtain the relation:
\begin{equation}
M_{core} = 0.17 \left(\frac{\Sigma_0}{\text{g~cm\textsuperscript{2}}}\right)^\frac{3}{2}\left(\frac{M_\star}{M_\odot}\right)^\frac{1}{2}\left(\frac{r}{\text{AU}}\right)^\frac{9}{4}\exp\left(\frac{r}{AU}\right)~\text{\earthmass}    
\end{equation}

To form $M_{core} \gtrsim 10$ \earthmass~, we need $r\gtrsim4.5$ AU, corresponding to an orbital period  $\gtrsim5400$ days. The formation timescale of the core with 10 \earthmass{} at this location can be estimated using Equation 20 of \cite{accr_timescale}, which can be written as:
\begin{equation}
  t_{pla}=1.7\text{Myr}~f_{pla}^{-1}\left(\frac{M}{10\text{M}_\oplus}\right)^{1/3}\left(\frac{r}{5 \text{AU}}\right)^{1.5} [max(\zeta,1)\zeta]  
\end{equation}

$f_{pla}$ is a parameterization of the surface density profile of the planetesimals, which is assumed to be 1 here \citep{refId01, accr_timescale}. $\zeta$  determines the accretion regime, which is assumed to be 1 for planetesimals in circular orbit \citep{Rafikov_2004, accr_timescale}. Therefore, the timescale to form the core of 10 \earthmass{} at $r=4.5$ AU is $\sim1.4$ Myr.

Since the planet likely formed \textit{ex-situ}, it would subsequently have to migrate to its present location. The estimated value of eccentricity ($\approx0$, see \autoref{tab:orbitalparam}), is consistent with the possibility of migration through disk interaction \citep{Baruteau_2014}. The timescale for type 1 migration, which is estimated using Equation 3 of \cite{Baruteau_2014} for TOI-5688 A b is $\gtrsim 1$ Myr from $\sim 4.5$ AU, for a disk with a gas-to-dust ratio of 100.

 Another possibility of planetary migration is through gravitational scattering \citep{Cloutier_2013}, followed by circularization of the planetary orbit \citep{Pont_2011}. However, using the equation provided by \cite{Pont_2011} with a tidal dissipation factor \( Q = 10^6 \), the circularization timescale is approximately \( \sim10^{12} \) years. As this timescale is greater than the age of the universe, and the lack of a highly eccentric orbit suggests that this planet is unlikely to have formed via gravitational scattering.
\subsubsection{Gravitational instability}
The formation of this planet by Gravitational Instability \citep[GI;][]{disc_inst1, disc_inst2} is hard to predict since it is affected by various factors such as protostellar disk mass, cooling prescription, etc. Although GI has been explored for the formation of GEMS \citep{2006ApJ...643..501B, 2023ApJ...956....4B}, TOI-5688 A b is at the lower end of the planet masses or mass ratios (planet-to-star mass ratio) typically seen as a result of GI \citep[etc.]{2003MNRAS.338..227R, 2006ApJ...643..501B, Boley_2009}. Studies by \cite{2006ApJ...643..501B, 2016ARA&A..54..271K} and others have shown the propensity of GI to form objects that are typically $\gtrsim 1$ M\textsubscript{J}. {Also, the simulations by \cite{Cai_2006} show that the strength of GI decreases as metallicity increases.} On the other hand, magneto-hydrodynamic simulations by \cite{2021NatAs...5..440D} show that disk fragmentation in the presence of magnetic fields in the disk could lead to the formation of intermediate-mass planets. Therefore, given a sufficiently massive protostellar disk \citep{2023ApJ...956....4B}, while GI is possible,  it is unnecessary to invoke GI as the necessary means of formation for this object.    

\subsection{Wide Separation Companion}

TOI-5688 A is a member of a wide-separation binary system (Section \ref{subsec:galactic}). \cite{el-bardy2021} have cataloged wide separation binary systems from Gaia DR3 with either the main sequence or white dwarf companion based on the location of the companion in the Gaia color-absolute magnitude diagram. It is estimated that $\sim40\%$ of M-dwarfs in the solar neighborhood have at least one companion within $\sim1000$ AU \citep{binary1, Lada_2006,binary3}. So far {10} (out of 25 transiting GEMS) are part of a binary system: HATS-74 \citep{Jordan_2022}, TOI-3984 \citep{Ca_as_2023}, TOI-5293 \citep{Ca_as_2023}, TOI-3714 \citep{Ca_as_2022}, K2-419 \citep{kanodia20246gems}, TOI-5634 \citep{kanodia20246gems}, TOI-6034 \citep{kanodia20246gems}, {TOI-762 A \citep{hartman2024toi762btic}, TOI-6383 \citep{bernabo2024searchinggemstoi6383abgiant}} and TOI-5688 (this work).

Multi-star systems are common in our galaxy, yet the influence of the companion star on the planetary system is still unclear. Studies have been conducted on the giant planet population that is orbiting FGK-type stars to test the significance of multi-body interactions on the planetary system \citep[eg:][]{Wang_2014, Knutson_2014, Ngo_2015, Evans_2018a}. Relative to field stars, giant planets with a period $<10$ days have been reported to have a high-wide binary fraction. \cite{fontanive2019} reports that $79^{+13.2}_{-14.7}\%$ of systems with a massive substellar object have a wide separation companion between 50 -- 2000 AU. However, \cite{moe_2021} revisited this assertion and concluded that the wide separation companions of hosts of giant planets in 50 -- 2000 AU do not significantly enhance the formation mechanism. Instead, the higher fraction of wide binaries among hot Jupiters is due to the inhibited formation of hot Jupiters in close binary systems. 

\section{Summary}\label{sec:summary}
In this paper, we present the discovery and characterization of a short-period Saturn-like exoplanet that is hosted by an M2V dwarf with T\textsubscript{eff} $=3713\pm59$~K. The planet has a mass $124.0^{+23.3}_{-24.4}$ \earthmass{} and radius of $10.4^{+0.6}_{-0.7}$ \earthradius. The density of this planet is $0.61^{+0.20}_{-0.15}$ g/cm\textsuperscript{3}, which makes it similar to Saturn.  The host is a member of a wide-separation binary system, with the companion being a main sequence star with an effective temperature of $3231^{+65}_{-62}$ K. W{e utilize  seven TESS sectors of photometry as well as ground-based transit follow-up with the 0.6 m RBO and 1.0 m TMF, as well as photometry from uTIRSPEC, speckle imaging from NESSI, and radial-velocity observations from HPF and NEID.}  The planetary and orbital parameters were estimated by Bayesian analysis with Hamiltonian Monte Carlo. The estimated stellar and planetary parameters support the core-accretion formation model for such Saturn-like exoplanets. 

\section*{Acknowledgements}
These findings stem from observations conducted using the Habitable-zone Planet Finder Spectrograph on the Hobby-Eberly Telescope (HET). We gratefully acknowledge support from various sources, including NSF grants AST-1006676, AST-1126413, AST-1310885, AST-1310875, AST-1910954, AST-1907622, AST-1909506, ATI 2009889, ATI-2009982, AST-2108512, AST-1907622, AST-2108801, AST-2108493 and the NASA Astrobiology Institute (NNA09DA76A), in our efforts to achieve precision radial velocities in the near-infrared (NIR). The HPF team is also appreciative of funding provided by the Heising – Simons Foundation through grant 2017-0494. Furthermore, we acknowledge the collaborative efforts of the University of Texas at Austin, the Pennsylvania State University, Ludwig-Maximilian-Universität München, and Georg-August Universität Gottingen in the Hobby–Eberly Telescope project. The HET is named in recognition of its principal benefactors, William P. Hobby and Robert E. Eberly.
The HET collaboration extends its appreciation to the Texas Advanced Computing Center for its support and resources. We express gratitude to the Resident Astronomers and Telescope Operators at the HET for their adept execution of observations with HPF. Additionally, we acknowledge that the HET is situated on Indigenous land. Furthermore, we wish to recognize and pay our respects to the Carrizo \& Comecrudo, Coahuiltecan, Caddo, Tonkawa, Comanche, Lipan Apache, Alabama-Coushatta, Kickapoo, Tigua Pueblo, and all the American Indian and Indigenous Peoples and communities who have inhabited or become part of these lands and territories in Texas, here on Turtle Island.
We acknowledge the Texas Advanced Computing Center (TACC) at The University of Texas at Austin for providing high-performance computing, visualization, and storage resources that have contributed to the results reported within this paper.
The data presented in this work were acquired at the WIYN Observatory using telescope time allocated to NN-EXPLORE through a scientific partnership involving the National Aeronautics and Space Administration (NASA), the National Science Foundation (NSF), and NOIRLab. Funding support for this research was provided by a NASA WIYN PI Data Award, which is managed by the NASA Exoplanet Science Institute. The observations were conducted with NEID on the WIYN 3.5 m telescope at Kitt Peak National Observatory (KPNO), under NSF's NOIRLab, and were carried out under proposal 2023A-633546 (PI: S. Kanodia), managed by the Association of Universities for Research in Astronomy (AURA) under a cooperative agreement with the NSF. We thank the NEID Queue Observers and WIYN Observing Associates for their skillful execution of our NEID observations. This work was performed for the Jet Propulsion Laboratory, California Institute of Technology, and sponsored by the United States Government under Prime Contract 80NM0018D0004 between Caltech and NASA. The WIYN Observatory is a collaborative effort involving the University of Wisconsin-Madison, Indiana University, NSF's NOIRLab, Pennsylvania State University, Purdue University, University of California-Irvine, and the University of Missouri. We acknowledge the privilege of conducting astronomical research on Iolkam Du’ag (Kitt Peak), a mountain of special significance to the Tohono O’odham people.
Several observations presented in this paper utilized the NN-EXPLORE Exoplanet and Stellar Speckle Imager (NESSI) under the proposal 2022B-936991. NESSI received funding from the NASA Exoplanet Exploration Program and the NASA Ames Research Center. Steve B. Howell, Nic Scott, Elliott P. Horch, and Emmett Quigley built NESSI at the Ames Research Center.

We utilized data from the Gaia mission\footnote{\url{https://www.cosmos.esa.int/gaia}} of European Space Agency (ESA), processed by the Gaia Data Processing and Analysis Consortium (DPAC, \footnote{\url{https://www.cosmos.esa.int/web/gaia/dpac/consortium}}). The DPAC's funding is provided by national institutions, particularly those involved in the Gaia Multilateral Agreement. Additionally, we acknowledge support from NSF grant AST-1907622 for conducting precise photometric observations from the ground.
We express gratitude for the assistance provided by NSF grant AST-1907622 in conducting meticulous photometric observations from the ground.
This study has utilized the Exoplanet Follow-up Observation Program website, managed by the California Institute of Technology under contract with the National Aeronautics and Space Administration as part of the Exoplanet Exploration Program.

This research has made use of the Exoplanet Follow-up Observation Program (ExoFOP; DOI: 10.26134/ExoFOP5) website, which is operated by the California Institute of Technology, under contract with the National Aeronautics and Space Administration under the Exoplanet Exploration Program.

VR, JPN, and DKO acknowledge the support of the Department of Atomic Energy, Government of India, under project identification No. RTI 4002.
VR and JPN were supported in part by a generous donation (from the Murty Trust) aimed at enabling advances in astrophysics through the use of machine learning. Murty Trust, an initiative of the Murty Foundation, is a not-for-profit organization dedicated to preserving and celebrating culture, science, and knowledge systems born out of India. Mrs. Sudha Murty and Mr. Rohan Murty head the Murty Trust.
CIC acknowledges support from NASA Headquarters through an appointment to the NASA Postdoctoral Program at the Goddard Space Flight Center, administered by ORAU through a contract with NASA.
{L.M.B acknowledges the support of the European Union’s Horizon Europe Framework Programme under the Marie Skłodowska-Curie Actions grant agreement No. 101086149 (EXOWORLD).}
VR would like to acknowledge the use of Grammarly, LanguageTool and ChatGPT for improving the English grammar and sentence structure in the manuscript.

We express our gratitude to the anonymous referee for providing valuable feedback, which has enhanced the overall quality of this manuscript.


\facilities{HET (HPF), WIYN (NEID, NESSI), \tess{}, RBO, TMF, HCT (TIRSPEC), Exoplanet Archive, Gaia, MAST}
\software{\texttt{ArviZ} \citep{arviz}, 
AstroImageJ \citep{astroimagej},
\texttt{astroquery} \citep{astroquery},
\texttt{astropy} \citep{astropy1,astropy2},
\texttt{barycorrpy} \citep{barycorrpy},
\texttt{celerite2} \citep{celerite21,celerite22},
\texttt{corner} \citep{corner},
ChatGPT \citep{openai2023gpt4},
\texttt{eleanor} \citep{eleanor},
Emacs \citep{emacs}
\texttt{EXOFASTv2} \citep{eastman2019exofastv2},
\texttt{exoplanet} \citep{exoplaneta},
\texttt{galpy} \citep{galpy_2015},
\texttt{HxRGproc} \citep{hxrgproc},
\texttt{HPF-SpecMatch} \citep{specmatch},
\texttt{ipython} \citep{ipython},
\texttt{lightkurve} \citep{lightkurve},
\texttt{matplotlib} \citep{matplotlib},
\texttt{numpy} \citep{numpy},
Org mode \citep{orgmode}
\texttt{pandas} \citep{pandas},
\texttt{photutils} \citep{larry_bradley_2024_10967176}
\texttt{pyastrotools} \citep{kanodia_2023_7685628},
\texttt{pyMC3} \citep{10.7717/peerj-cs.55}, 
\texttt{scipy} \citep{scipy},
\texttt{SERVAL} \citep{serval},
\texttt{starry} \citep{2018ascl.soft10005L},
\texttt{telfit} \citep{2014AJ....148...53G}
\texttt{Theano} \citep{theano}
}

\bibliography{References}

\end{document}